\documentclass[10pt,twocolumn,english]{article}
\usepackage{lmodern}

\usepackage[T1]{fontenc}
\usepackage[latin9]{inputenc}
\usepackage{babel}
\usepackage{verbatim}
\usepackage{varioref}
\usepackage{amsmath}
\usepackage{amssymb}

\makeatletter
\providecommand*{\code}[1]{\texttt{#1}}

\@ifundefined{date}{}{\date{}}
\makeatother

\begin{document}
\title{Java Generics: An Order-Theoretic Approach}
\author{Moez A. AbdelGawad\\
Informatics Research Institute, SRTA-City, Alexandria, Egypt\\
\texttt{moez@cs.rice.edu}}
\maketitle
\begin{abstract}
Generics have been added to Java so as to increase the expressiveness
of its type system. Generics in Java, however, include some features---such
as Java wildcards, $F$-bounded generics, and Java erasure---that
have been hard to analyze and reason about so far, reflecting the
fact that the mathematical modeling of generics in Java and other
similar nominally-typed object-oriented programming (OOP) languages
is a challenge. As a result, the type systems of mainstream nominally-typed
OOP languages, which are built based on current models of generics,
are overly complex, which hinders the progress of these type systems.

In this paper we present a detailed outline of a new approach to modeling
Java generics that uses concepts and tools from order theory, and
we report on our progress in developing this approach. Fundamentally,
we use the nominal subclassing relation (as a partial order) together
with some standard and novel order-theoretic tools to construct the
generic nominal subtyping relation (as a partial order) and the containment
relation between generic type arguments (a third partial order). We
further analyze the relation between these three ordering relations---which
lie at the heart of mainstream generic OO type systems---using order
theoretic tools, and accordingly we explore extensions of OO type
systems suggested by such analysis. %
In our approach we also make use of some concepts and tools from category
theory. We believe a combined order-theoretic and category-theoretic
approach to modeling generics holds the keys to overcoming much of
the adversity found when analyzing features of generic OO type systems.
\end{abstract}

\paragraph{Introduction}

The addition of generic classes, generic interfaces and parameterized
types (\emph{i.e.}, instantiations of generic classes\footnote{In this paper interfaces are treated as abstract classes. The term
`classes' here thus refers to Java classes and Java interfaces.
Also, parameterized types are sometimes called \emph{reference types},
\emph{object types}, \emph{class types}, \emph{generic types}, or
just \emph{types}.}) to Java has significantly enhanced the expressiveness of its type
system~\cite{JLS05,JLS18,Bracha98,Corky98,Thorup99,GenericsFAQWebsite}.
Support for generics in Java and other mainstream nominally-typed
OOP languages similar to it---such as C\#~\cite{CSharp2015}, Scala~\cite{Odersky14},
C++~\cite{CPP2017}, and Kotlin~\cite{Kotlin18}---has several
drawbacks however.

For example, the bounded existentials model of Java wildcards, while
accurate, is exceedingly complex~\cite{Torgersen2004,MadsTorgersen2005,Cameron2007,Cameron2008,Summers2010,Tate2011}.
The notion of ``capture conversion,'' which is based on the bounded
existentials model of wildcards, renders the generic type system of
Java fully comprehensible only to the most advanced of Java developers\footnote{Check, for example, sections of the JLS---the Java Language Specification---that
specify crucial parts of its generic type system, \emph{e.g.}, \cite[Sec 4.5 \& Sec. 5.1.10]{JLS18}.}.

Further, support of some features in Java generics has a number of
irregularities or ``rough edges.'' These include type variables
that can have upper $F$-bounds but cannot have lower bounds (let
alone lower $F$-bounds), wildcard type arguments that can be upper-bounded
or lower-bounded but not both, and Java erasure---a feature particular
to Java and Java-based OO languages such as Scala and Kotlin---that
is usually understood, basically, as being ``outside the type system.''

In this paper we present the outline of a model of Java generics---one
based on order theory and category theory---that seemingly promises
to help in overcoming many if not most of the difficulties met when
modeling generics using current approaches. Our approach uses concepts
from order theory such as products of posets, intervals over posets,
and in\-duc\-tive/co\-induc\-tive mathematical objects%
, and it also uses in an elementary way some simple tools and concepts
from category theory such as algebras, coalgebras, adjunctions, monads
and operads.

As an application that demonstrates its value, our approach enables
us to suggest how the type system of Java and similar nominally-typed
OOP languages can be simultaneously streamlined and extended to support
new features such as interval types, lower $F$-bounds, and co-free
types. As a side-benefit, the approach also allows us to develop some
mathematical vocabulary (such as $F$-subtypes, $F$-supertypes, and
free types) for referring to some notions in generic nominally-typed
OO type systems that have been mathematically unidentified so far.

\paragraph{Constructing The Generic Java Subtyping Relation}

Nominally-typed OOP languages such as Java, C\#, C++, Scala and Kotlin
are class-based. Two fundamental ordering relations in statically-typed
class-based OOP languages are the \emph{subclassing }(also called
\emph{inheritance}) relation between classes and the \emph{subtyping}
relation between reference types.

The first step in our order-theoretic approach to Java generics is
defining an abstract partial product operation on posets (\emph{i.e.},
ordered sets), which we call \emph{ppp} (for partial poset product)
and denote by $\ltimes$. The \emph{ppp }operation constructs a product-like
output poset given two input posets and a subset of the first input
poset.

Intuitively, the \emph{ppp} operation on two posets simply pairs \emph{some}
elements of the first input poset with all elements of the second
input poset, then it adds the unpaired elements of the first poset
to the resulting product. The \emph{ppp} operation then orders the
resulting elements (\emph{i.e}., pairs and non-pairs) based on the
orderings in the two input posets.\footnote{For comparison, a standard (also called \emph{tensor} or \emph{direct})
product of two posets pairs \emph{all} elements of the first poset
with all those of the second, then orders the resulting pairs based
on the orderings in the two input posets~\cite{Davey2002} (which
is the same as the tensor product---sometimes, imprecisely, also
called the Cartesian product---of the two directed graphs corresponding
to Hasse diagrams of the two input posets~\cite[Ch. 32]{Hammack2011}).} The formal definition of \emph{ppp }is the order-theoretic counterpart
of the definition of partial Cartesian product of graphs presented
in~\cite{AbdelGawad2018a}.

Second, we define a unary \emph{wildcards} (also called \emph{triangle})
operation on posets, which we abbreviate \emph{wc} and denote by $\triangle$.
The \emph{wc }operation produces as output a (roughly triangle-shaped)
poset out of an input \emph{bounded} poset (\emph{i.e.}, a poset with
top and bottom elements).

Intuitively, the \emph{wc }operation constructs three elements (modeling
three type arguments) in the output poset corresponding to each element
(a type) in the input poset. The three output elements are two wildcard
type arguments and one non-wildcard type argument. These three elements
are then ordered (by a relation called \emph{containment}) based on
the ordering in the input poset (the subtyping relation). Namely,
for an element (\emph{i.e.}, a type) \code{T} in the input poset,
the \emph{wc }operation constructs the three elements (\emph{i.e.},
three type arguments) `\code{?~<:~T}' (a synonym for the Java wildcard
type argument `\code{?~extends~T}'), `\code{T~<:~?}' (a synonym
for `\code{?~super~T}'), and the non-wildcard type argument `\code{T}',
then it orders these three elements by containment (where the non-wildcard
type argument `\code{T}' is \emph{contained-in} both wildcard type
arguments `\code{?~<:~T}' and `\code{T~<:~?}'). The formal definition
of the wildcards operation \emph{wc} is presented in~\cite{AbdelGawad2018b}.

In OO programming languages, the subtyping relation between ground
types (\emph{i.e.}, ones with no type variables) is the basis for
the full generic subtyping relation (\emph{i.e.}, with type variables)~\cite{TAPL}.
Under some simple simplifying assumptions, the subclassing (\emph{i.e.},
inheritance) relation between Java classes can be used to construct
the subtyping relation between ground generic Java types, iteratively,
using the \emph{ppp} and \emph{wc} operations.

In particular, the subtyping relation is constructed, iteratively,
as a the result of the \emph{ppp} of two posets (relative to a subset
of the first), where
\begin{itemize}
\item The first input poset to the \emph{ppp }($\ltimes$) operation is
that of the \emph{subclassing} relation (\emph{i.e.}, of classes ordered
by inheritance),
\item The subset of the first poset (which the \emph{ppp} operation is defined
relative to) corresponds to the subset of \emph{generic} classes (which
is always a subset of the first subclassing poset), and
\item The second input poset to the \emph{ppp }operation is that of type
arguments (ordered by containment), which is produced by applying
the unary \emph{wc} ($\triangle$) operation to the subtyping relation
produced by the previous iteration in the construction process.
\end{itemize}
The iterative process of constructing the subtyping relation between
ground generic types in Java was first presented in~\cite{AbdelGawad2017a}.
And a summary of the iterative construction process together with
recursive poset equations that formalize the process---and that use
the $\ltimes$ and $\triangle$ operators---are presented in~\cite{AbdelGawad2018b}.\footnote{Formally, the poset of the subtyping relation $S$ is the (least)
solution of the recursive poset equation
\begin{equation}
S=C\ltimes_{C_{g}}\triangle\left(S\right)\label{eq:SC}
\end{equation}
where $C$ is the subclassing poset and $C_{g}$ is the subset of
generic classes of $C$.

Intuitively, Equation~(\ref{eq:SC}) can be read as saying that the
$\triangle$ operator constructs wildcard type arguments (and the
associated containment relation) from the subtyping relation $S$,
then, while preserving existing subtyping relations, the $\ltimes$
operator pairs the constructed type arguments with generic classes
in $C$ to construct further types and further subtyping relations.}

\paragraph{Extending Java Generics and Extending Its Order-Theoretic Model}

As we just discussed, the order-theoretic approach to Java generics
led to constructing the basis of the subtyping relation in Java using
order-theoretic tools. We now illustrate the extrapolating value of
the approach by suggesting two extensions of Java generics that are
inspired by this approach, as well as offering new order-theoretic
concepts and tools that can be useful in analyzing generic OO type
systems.

\label{introtext}In particular, we first suggest how wildcard types
can be generalized to what we call \emph{interval types}. Then we
suggest how $F$-bounded generics (\emph{a.k.a.}, $F$-bounded polymorphism)
can be generalized to what we call \emph{doubly $F$-bounded generics.}
Both extensions are suggested by the construction of the subtyping
relation between ground generic Java types we presented above.

In\-duction and co\-induc\-tion are mathematical concepts that
are best studied %
in the context of order theory and category theory. As such, during
our discussion of doubly $F$-bounded generics in particular, the
order-theoretic approach naturally leads us to consider viewing generic
OO classes as generators (of types) over the poset of types ordered
by the nominal subtyping relation (\emph{i.e.}, the subtyping relation
between ground generic Java types), and, thus, to involve (co)induc\-tive
types and \emph{mutual} (co)induc\-tive types in the discussion.

\paragraph*{{*}Interval Types}

Intuitively, interval types generalize wildcard types by generalizing
wildcard type arguments to interval type arguments. Interval type
arguments are type arguments that can have \emph{any} two types as
the upper and lower bounds of the type argument, as long as the lower
bound is a subtype of the upper bound (so as to define, in an order-theoretic
sense, an interval over the poset of the subtyping relation).

Formally, with interval types, the Java subtyping relation can be
constructed simply by substituting the wildcards operation \emph{wc}
we discussed above with a unary operation that constructs intervals
of an input poset. We call this operation on posets \emph{int}, and
denote it by $\Updownarrow$.\footnote{Formally, the poset of the subtyping relation $S$ (with interval
types) is the (least) solution of the recursive poset equation
\begin{equation}
S=C\ltimes_{C_{g}}\Updownarrow\left(S\right)\label{eq:SC-1}
\end{equation}
where $C$ is the subclassing poset and $C_{g}$ is the subset of
generic classes of $C$.

Intuitively, Equation~(\ref{eq:SC-1}) can be read as saying that
the $\Updownarrow$ operator constructs interval type arguments (and
the associated containment relation) from the subtyping relation $S$,
then, while preserving existing subtyping relations, the $\ltimes$
operator pairs the constructed type arguments with generic classes
in $C$ to construct further types and further subtyping relations.}

Intuitively, the unary \emph{int }operation constructs an element
(an interval type argument) in the output poset corresponding to each
\emph{pair} of elements (types) in the input (subtyping) poset where
the first component of the pair is less than or equal (\emph{i.e.},
is a subtype) of the second component. All constructed elements are
then ordered (also by the \emph{containment} relation) based on the
ordering in the input poset (\emph{i.e.}, the subtyping relation).
Namely, for a pair of elements (\emph{i.e.}, types) \code{S} and
\code{T} in the input poset where \code{S<:T} (\emph{i.e.}, \code{S}
is a subtype of \code{T}, thus defining an interval in the input
poset), the \emph{int }operation constructs the element `\code{{[}S,T{]}}',
then it orders all such constructed elements by containment (\emph{i.e.},
in the output poset the interval type argument `\code{{[}S,T{]}}'
is \emph{contained-in} $\left(\sqsubseteq\right)$ the interval type
argument `\code{{[}U,V{]}}' if and only if \code{U<:S} \emph{and}
\code{T<:V} in the input poset).\footnote{The \emph{int} operation typically produces \emph{more} type arguments
than the \emph{wc} operation does, since \emph{int} produces a set
of type arguments corresponding to all \emph{pairs} of types in the
subtyping relation where the first type is a subtype of the second
(which, due to the existence of types \code{Object} and \code{Null}
at the top and bottom of the subtyping relation, contains all wildcard
type arguments as a subset). (N.B.: If $n=\left|P\right|$ is the
cardinality of an input poset $P$, then\emph{ $\left|wc\left(P\right)\right|=3n-2$
}while $\left|int\left(P\right)\right|=O\left(n^{2}/2\right)$, \emph{i.e.},
is on the order of $n^{2}/2$.)

Similar to the \emph{wc }operation, the \emph{int }operation orders
interval type arguments by the containment relation (corresponding
to the containment relation between intervals of a poset or between
paths of a graph). As such, interval types are strictly more expressive
than wildcard types (if there is at least one generic class in the
subclassing relation, and is equally expressive otherwise).} The formal definition of \emph{int }is presented in~\cite{AbdelGawad2018c}.

Defining interval types allows immediately noting that, when viewed
as functions (\emph{i.e.}, generators, or type constructors), generic
classes are \emph{not} endofunctions (\emph{i.e.}, self-maps), since
they do not map types to types, but they rather map type intervals
to types (\emph{i.e.}, map interval type arguments to interval types).
Further, we also note that generic classes are always \emph{monotonic}
functions (also called \emph{covariant} \emph{functors} in category
theory parlance). This is because generic classes, as generators of
types from interval type arguments, always produce \emph{subtypes}
when they are provided with \emph{subintervals} as type arguments.
See~\cite{AbdelGawad2019b} for more details.

\paragraph*{{*}Doubly $F$-Bounded Generics}

Motivated by generalizing wildcard types to interval types, we next
define doubly $F$-bounded generics (\emph{dfbg}, for short)\emph{
}as a generalization of standard $F$-bounded generics, where, in
\emph{dfbg}, type variables of generic classes---which are allowed
to have upper $F$-bounds in Java---are allowed to also have \emph{lower}
bounds, including lower $F$-bounds. Our investigations into \emph{dfbg}
led us, among other conclusions, into making a distinction between
\emph{valid} type arguments and \emph{admittable} type arguments,
and between \emph{valid} parameterized types and \emph{admittable}
parameterized types.

To illustrate with an example, type \code{Enum<Object>} is an admittable
Java parameterized type, but it is not a valid type since the type
argument to \code{Enum}, namely \code{Object}, does not satisfy
the declared bounds on the type variable of class \code{Enum}. Hence,
type \code{Object} is an admittable type argument to generic class
\code{Enum}, but it is not a valid one.

This example illustrates the simple definition of admittable versus
valid parameterized types and type arguments. All reference types
are \emph{admittable} type arguments. An admittable type argument
passed to a generic class (to form a parameterized type) is \emph{valid}
if and only if it satisfies the bound(s) of the corresponding type
variable in the class (\emph{i.e.} is a subtype of the upper bound
and a supertype of the lower bound). An \emph{admittable} parameterized
type is formed by a generic class instantiated with admittable type
arguments. An admittable parameterized type is then a \emph{valid}
parameterized type if and only if \emph{all} its type arguments are
valid type arguments.

In Java, an intuitive, but somewhat inaccurate, set-theoretic way
to think about \emph{invalid} types (\emph{i.e.}, admittable-but-not-valid
reference types, such as \code{Enum<Object>}) versus valid types
(such as \code{List<String>}) is to consider valid types as denoting
\emph{non-empty} sets of objects (\emph{i.e.}, sets, that are, at
the very least, inhabited by the trivial object \code{null} as well
as ``the bottom object'' $\bot$, corresponding to computational
divergence~\cite{NOOPsumm}, in addition to possibly many other proper
non-trivial objects), while considering invalid types as denoting
the empty set (of objects), \emph{i.e.}, the uninhabited set that
does not contain \code{null}, nor even $\bot$. In this view, thus,
admittable Java types correspond to all (\emph{i.e.}, to empty or
non-empty) sets of objects, while all invalid Java types are different
type expressions that, in spite of their (syntactic) differences,
semantically denote the same empty set of objects.\footnote{For the sake of completeness, examples of Java type expressions that
are \emph{not} admittable (reference) types (let alone valid types)
are the type expressions \code{int} and \code{boolean}, since \code{int}
and \code{boolean} are not reference types, and also type expressions
similar to \code{String<Object>}, since class \code{String} is not
a generic class and thus cannot be applied to or instantiated with
any type arguments. (Like all zero-ary type constructors, which take
no type arguments, class \code{String} constructs only one type,
namely the homonymous type \code{String}.)} More details on doubly $F$-bounded generics are presented in~\cite{AbdelGawad2018e}%
.

It is worthy to note that, inspired by functions in analysis (\emph{i.e.},
functions over real numbers $\mathbb{R}$), in the investigation of
\emph{dfbg} we (attempt to) use a \emph{coinductive} logical argument
to prove%
{} that checking the validity of some type arguments inside the bounds-declarations
of generic classes is \emph{unnecessary}. As such, we conclude that
using \emph{admittable} type arguments in such a context is allowed.
See~\cite{AbdelGawad2018e} for details.

\paragraph*{{*}(Co)Induc\-tive Types, $F$-Sub\-types, $F$-Super\-types,
and Mutual (Co)In\-duction}

In logic, coinductive reasoning can, intuitively, be summarized as
asserting that a statement is proven to be true if there is \emph{no}
(finite or ``good'') reason for the statement to \emph{not} hold~\cite{Kozen2016,AbdelGawad2019a}.
Such a statement is then said to be coinductively-justified, or `true
by coinduction'. Showing the value of the order-theoretic approach
yet again, order theory (which lattice theory is a subdiscipline of)
is regarded as the most natural context for studying inductive and
coinductive mathematical objects and the associated logical proof
principles (see, \emph{e.g.},~\cite[Ch. 21]{TAPL}).\footnote{See~\cite{AbdelGawad2018f} for a more in depth discussion and comparison
of formulations of induction and coinduction in different mathematical
disciplines.}

Among a few other motivating factors, Tate et al.'s conclusion that
Java wildcards are \emph{coinductive }bounded existentials~\cite{Tate2011}
and the use of a coinductive argument during investigating \emph{dfbg},
led us to make note of and consider, in more depth and in more generality,
the coinductive nature of nominally-typed OO type systems~\cite{AbdelGawad2019b}.
Another motivation for studying coinduction is to allow for analyzing
the subtyping relation in Java from a coinductive point of view, so
as to allow, for example, for subtyping relations that are \emph{not}
constructed finitely~\cite{AbdelGawad2019b}.

In the course of studying induction and coinduction, we define new
notions such as $F$-subtypes and $F$-supertypes and discuss their
relevance to OO type systems~\cite{AbdelGawad2019b}. In summary,
if $F$ is a generic class, then the $F$-subtypes of class $F$ are
all the parameterized types that are subtypes of instantiations of
$F$ with themselves (\emph{i.e.}, any reference type \code{Ty} is
an $F$-subtype of class $F$ if and only if \code{Ty~<:~F<Ty>}).
Dually, the $F$-supertypes of $F$ are all the parameterized types
that are supertypes of instantiations of $F$ with themselves (As
such, type \code{Object} in Java, for example, is an $F$-supertype
while type \code{Null} is an $F$-subtype for all generic classes
$F$.)

In our experience, $F$-subtypes and $F$-supertypes are useful, mainly,
when discussing and analyzing bounded type variables in generic OO
type systems (\emph{e.g.}, as we do for \emph{dfbg}). The two notions,
for example, establish a direct relation and correspondence between
the denoted types and $F$-al\-gebras and $F$-co\-algebras (sometimes
also called just algebras and coalgebras) in category theory, and
also between the denoted types and in\-ductive/co\-induc\-tive
mathematical objects\emph{ }(\emph{i.e.}, sets, points, predicates,
and structural types) in each of set theory, order theory, first-order
logic, and structural type theory (which is used to model \emph{functional}
programming languages)~\cite[Tab. 1 \& 2]{AbdelGawad2018f}.\footnote{In order theory in particular, $F$-supertypes correspond to \emph{pre-fixed}
points, while $F$-subtypes correspond to \emph{post-fixed} points.
In set theory and structural type theory, they correspond to \emph{inductive}
sets/types and \emph{coinductive} sets/types, respectively.}

Noting that classes in Java programs, including generic classes, are
frequently defined mu\-tually-recur\-sively (\emph{e.g.}, assuming
the absence of primitive types in Java, the definitions of the fundamental
classes \code{Object} and \code{Boolean} are mutually dependent
on each other), we were also led to define an order-theoretic notion
of \emph{mutual coinduction} to enable studying mutually recursive
definitions~\cite{AbdelGawad2019}. Mutually-dependent definitions
are not only frequent in OO programs (\emph{i.e.}, in defining methods
and classes), but they show up also in the definition of OO programming
languages themselves (\emph{e.g.}, as in the definition of the Java
subtyping relation, where the subtyping relation between parameterized
types depends on the containment relation between interval type arguments,
and, reciprocally, the containment relation between interval type
arguments depends on the subtyping relation between parameterized
types. Also, there exists a mutual dependency between the definitions
of valid and admittable types and type arguments). We believe the
order-theoretic notion of mutual coinduction, as we define it, can
be useful in analyzing OO type systems and also in reasoning about
OO software (and imperative software). See~\cite{AbdelGawad2019}
for further motivations and details on mutual coinduction.

\paragraph{Category Theory}

Given that, in a precise sense, category theory can be viewed as a
(major) generalization of order theory~\cite{Fong2018,Priestley2002,spivak2014category},
category theory can enter into our order-theoretic approach to modeling
Java generics via using at least three category-theoretic tools:

\paragraph*{{*}Adjunctions}

In the order-theoretic approach to modeling Java generics, a clear
distinction is made and maintained between classes and subclassing,
on one hand, and types and subtyping, on the other hand.%

This clear distinction between classes, as type constructors, and
types, as constructed by classes, allows us to easily see that an
adjunction~\cite{spivak2014category} (called a `Galois connection'
in order theory parlance~\cite{Davey2002}) exists between subclassing,
as a relation between generic and non-generic classes, and subtyping,
as a relation between parameterized types. We call this adjunction
the Java Erasure Adjunction (\emph{JEA}).

In \emph{JEA}, Java erasure, which ``erases'' the type arguments
of a parameterized type, is the left adjoint. The notion of a \emph{free
type},\emph{ }which maps a class to a type expressing the ``most
general wildcard instantiation'' of the class, is the right adjoint
of \emph{JEA}.

As the left part (or adjoint) of the adjunction, Java erasure maps
a parameterized type to a class (by ``erasing'' the type arguments),
while as the right part of the adjunction, the notion of a \emph{free
type} corresponding to a class maps any generic class to the type
expressing the ``most general wildcard instantiation'' of the class
(\emph{e.g.}, a generic class \code{C} with one type parameter is
mapped to the type \code{C<?>} as its corresponding free type). We
call this adjunction the \emph{Java Erasure Adjunction} (\emph{JEA}).

As for any adjunction, to properly define an adjunction the two maps
of \emph{JEA} (from the subclassing poset to the subtyping poset,
and vice versa) have to work in tandem to satisfy a preservation condition.
This condition indeed holds for Java generics, making Java erasure
and free types two adjoints (parts) of an adjunction, hence deserving
their names as adjoints in an adjunction. In particular, if $E$ is
the erasure mapping and $FT$ is the free type mapping, and if $\le$
denotes the subclassing relation and $<:$ denotes the subtyping relation,
then for $E$ and $FT$ to define an adjunction it should be the case
that 
\begin{equation}
E(t)\le c\Longleftrightarrow t<:FT(c),\label{eq:adj}
\end{equation}
for all types $t$ and classes $c$.

In words, this condition says that the erasure $E(t)$ of a parameterized
type $t$ is a subclass of class $c$ if and only if $t$ is a subtype
of the free type $FT(c)$ corresponding to class $c$.\footnote{Consider, for example, the statement (in Java) 
\[
\mathtt{LinkedList\le List\Longleftrightarrow LinkedList\negthickspace<\negthickspace T\negthickspace>\;<:\;List\negthickspace<?\negthickspace>}
\]
where, in Equation~(\ref{eq:adj})\vpageref{eq:adj}, type variable
$t$ is instantiated to the generic type \code{LinkedList<T>} for
all type arguments $\mathtt{T}$ (\emph{e.g.}, \code{String} or \code{Integer}
or \code{?~extends Number}) and class variable $c$ is instantiated
to class \code{List}. This statement asserts that class \code{LinkedList}
in Java is a subclass of \code{List} if and only if all instantiations
of \code{LinkedList} are subtypes of the free type \code{List<?>}---which
is a true statement in Java.}

The preservation condition expressed by Equation~(\ref{eq:adj})\vpageref{eq:adj}
is equivalent to the statement stating that, for any two classes \code{C}
and \code{D}, if \code{D} is a subclass of (\emph{i.e.}, inherits
from) \code{C} then all parameterized types that are instantiations
of \code{D}, and their subtypes, are subtypes of the free type \code{C<?>}
corresponding to class \code{C} \emph{and} vice versa, \emph{i.e.},
if all instantiations of some class \code{D} and their subtypes are
subtypes of the free type \code{C<?>} corresponding to some class
\code{C}, then \code{D} \emph{is }a subclass of \code{C}.

As stated here, this statement is familiar to OO developers using
nominally-typed OO programming languages such as Java, C\#, C++, Scala
and Kotlin. It is a true statement in these languages due to the nominality
of subtyping in these languages. Subclassing (\emph{a.k.a.}, inheritance),
a relation characteristic of class-based OOP, is always specified
between classes in OO programs using class \emph{names}. Nominal subtyping
asserts a bidirectional correspondence between the subtyping relation
and the inherently nominal subclassing relation.

In the case of \emph{non}-generic OOP, the correspondence between
subtyping and subclassing is expressed, succinctly, by stating that
`inheritance \emph{is} subtyping'.. In the case of generic OOP,
the correspondence is succinctly expressed by stating that `inheritance
is \emph{the} source of subtyping'. The latter is a compact expression
of Equation~(\ref{eq:adj}).

Focusing on generic nominally-typed OOP, the `inheritance is the
source of subtyping' statement and Equation~(\ref{eq:adj}) state,
first (in the left-to-right direction), that subclassing does result
in (\emph{i.e.}, is \emph{a} source of) subtyping between reference
types, and, secondly (in the right-to-left direction), that subclassing
is the \emph{only }source of subtyping between reference types (\emph{i.e.},
that besides subclassing there are no other sources for subtyping).

It should be noted that the notion of the free type corresponding
to a class is similar, in a precise category-theoretic sense, to the
notion of the free monoid corresponding to a set and of the free category
(a quiver) corresponding to a graph. More details on free types and
\emph{JEA} are available in~\cite{AbdelGawad2017b},~\cite{AbdelGawad2019h}
and~\cite{AbdelGawad2019b}.

\paragraph*{{*}Monads}

Closure and kernel operators in order theory correspond to monads
and comonads in category theory~\cite{Fong2018,Priestley2002}. As
such, the discussion of order-theoretic (co)induc\-tion we presented
earlier (which can be expressed using closure and kernel operators~\cite{Davey2002})
can be generalized to involve categories%
, using monads and comonads.\footnote{Generalizing the discussion of in\-duction/co\-induc\-tion to categories
has the benefit of allowing the discussion of inductive and coinductive
types while \emph{not} requiring the subtyping relation to be a complete
lattice. This makes category theory and order theory (of non-lattices)
closer and better suited to study nominally-typed OO type systems,
since, by not requiring least upper bounds (lubs) and greatest lower
bounds (glbs), the two disciplines allow but do \emph{not} require
the existence of fixed points (of generators, \emph{i.e.}, type constructors),
unlike the case of structural type systems that are modeled by (power)
set theory. We further discuss this point in~\cite{AbdelGawad2018f,AbdelGawad2019b}.} When using monads from category theory, the discussion of inductive
types and coinductive types can be expressed, more generally, using
the category-theoretic notions of $F$\nobreakdash-algebras and $F$\nobreakdash-coalgebras
(or, algebraic types and coalgebraic types).

This observation allows us to easily see that $F$-subtypes and $F$-supertypes
of a generic class $F$ in a Java program correspond, in a precise
category-theoretic sense, to coalgebras and algebras of class $F$,
respectively. This further allows easily seeing that free types (as
the greatest/largest $F$-subtypes) are \emph{final} coalgebras in
the Java subtyping category (\emph{i.e.}, when the Java subtyping
relation is viewed as a category, rather than a poset), and that,
on the other hand, \emph{initial} algebras rarely exist in the Java
subtyping relation (since, unlike for free types, Java does \emph{not}
define a general notion of types that correspond to least/smallest
$F$-supertypes).

Discussing final coalgebras and initial algebras motivates us to also
suggest adding \emph{co-free types} as the least $F$-supertypes,
\emph{i.e.}, as initial algebras, to Java. As the name indicates,
cofree types function as duals of free types, which, as we discussed,
are final coalgebras in the Java subtyping relation.\footnote{In homage to Java's mascot, we sometimes call cofree types ``coffee
types!''.}

We did not investigate co-free types in much depth. However, for each
generic class \code{C}, we suggest the notation \code{C<!>} for
the corresponding co-free type. In Java the cofree type \code{C<!>}
has as its \emph{only} instance the trivial object \code{null} (with
the type adding knowledge that this object is of type \code{C<!>},
and thus can be used as an instance of class \code{C}, which is information
that type \code{Null} does not always tell, \emph{e.g.}, in lower
bounds). Further, the cofree type \code{C<!>} has as its supertypes
all parameterized types that are instantiations of class \code{C}
(and their supertypes), and has as its subtypes only the cofree types
corresponding to all subclasses of \code{C} (including type \code{Null}).
As such, just like the Java subtyping relation between free types
alone (\emph{i.e.}, when the relation is restricted to these types
only) being the same as (\emph{i.e.,} is order-isomorphic to) the
subclassing relation, the subtyping relation between co-free types
alone is also the same as the subclassing relation.

We envision the main use of cofree types to be as lower bounds of
type variables (in doubly $F$-bounded generics) and of interval type
arguments (in interval types), rather than, confusingly, using free
types. (In Java, currently, when a free type such as \code{C<?>}
is used as the lower bound of a wildcard type argument---as in \code{?~super~C<?>}---it
has a somewhat \emph{different }meaning than when the free type is
used as an upper bound of a wildcard type argument---as in \code{?~extends~C<?>}---hence
hinting at the need for co-free types. See~\cite{AbdelGawad2018e}
and~\cite{AbdelGawad2019h} for more details. We envision \code{?~super~C<!>}
as expressing more closely the intended meaning of \code{?~super~C<?>}.)

\paragraph*{{*}Operads}

In category theory, operads are a useful tool for modeling self-similar
phenomena~\cite{spivak2014category}. As such, operads can be used
to construct the generic subtyping relation in Java. The construction
process of the subtyping relation between ground Java types we discussed
earlier is an iterative process, reflecting the self-similarity of
the relation. As such, an operad can be defined to model this construction
process. In~\cite{AbdelGawad2017a} we presented an outline for defining
such an operad.

\paragraph{Discussion}

In this paper we presented the outline of an order-theoretic approach
to modeling Java generics. Unlike many extant models of gene\-ric
OOP, the order-theoretic approach fully embraces nominal typ\-ing/sub\-typ\-ing,
as found in mainstream OOP languages such as Java, C\#, C++, Scala
and Kotlin.

For example, in agreement with the nominality of the subtyping relation
in nominally-typed OOP, the nominal type in\-herit\-ance relation
is the \emph{sole} basis for defining the subtyping relation in our
approach.

Also, influenced by paths in graphs (which always have start and end
points) and by bounded posets (which always have top and bottom elements),
in our approach both \emph{upper} bounds and \emph{lower} bounds of
type variables and of type arguments are treated, in the same way,
as bounds of an interval.

For example, the order-theoretic model of Java generics includes an
explicit type \code{Null} (in symmetry with type \code{Object}),
and it suggests interval types as the proper double (\emph{i.e.},
upper and lower) bounded generalizations of wildcard types (\emph{e.g.},
\code{Null} is paired with \code{Object} to define the largest type
interval), and the model also suggests doubly $F$-bounded generics
as the proper generalization of $F$-bounded generics.\footnote{Further, similar to how vertices in graphs are sometimes treated as
trivial (`zero-length') paths (or trivial graph intervals) when
necessary, single types in the order-theoretic model are also sometimes
treated as `trivial interval types'.}

We believe the order-theoretic approach to modeling Java generics,
when developed in full, can offer an accurate model of generics (given
the few simplifying assumptions it assumes). The current approach
to modeling generics is based on bounded existentials, which is a
notion that is typically beyond the reach of most average Java developers
(and even, sometimes, advanced Java developers). Given its ultimate
dependency on \emph{finite} graphs and posets of the finite nominal
subclassing relation, we believe the order-theoretic approach is simpler
and more intuitive than the bounded existentials approach.

We believe that the use of bounded existentials as a model of Java
wildcards is due to most of earlier approaches to modeling generics
\emph{not} having nominal typ\-ing and sub\-typ\-ing as fundamental
characteristics of these approaches. Given the vast amount of earlier
research done on \emph{functional} programming languages, which are
largely \emph{structurally-typed}, most of these earlier approaches
to modeling generics were developed, at least initially, with structural
typ\-ing and sub\-typ\-ing in mind. This led these earlier approaches
to also adopt the bounded existentials model (of bounded parametric
polymorphism in functional programming~\cite{MilnerPolymorphism78})
as a model for Java wildcards. However, the structural and \emph{set-theoretic}
(rather than nominal and order-/cat\-egory-theo\-retic) orientation
of bounded parametric polymorphism, and thus of bounded existentials,
makes them more suited for modeling polymorphic structurally-typed
OOP languages (such as ML) but lesser suited for modeling generic
nominally-typed OOP languages (such as Java, C\#, Scala, Kotlin, and
C++).\footnote{As we discuss in more detail in~\cite{AbdelGawad2019b}, the structural
versus nominal orientation of each approach reflects itself, for example,
in set-theoretic models (like in structural type theory) necessitating
the existence of (least and greatest) fixed points of type gener\-ators/cons\-truc\-tors
as models of inductive and coinductive types, while order-theoretic
and category-theoretic models (like in nominal type theory) allow
the existence of such fixed points, but they do \emph{not} necessitate
their existence to model inductive and coinductive types (since these
types can be modeled instead by least \emph{pre}-fixed points and
greatest \emph{post}-fixed points in non-lattices order theory, or
by initial algebras and final coalgebras in category theory).} In other words, we believe having nominal typ\-ing and nominal
sub\-typ\-ing, and their immediate type-theoretic consequences,
at the heart of modeling Java generics allows having a model of Java
generics that is simpler than extant models. We hope this paper has
demonstrated so.\footnote{\emph{Collaboration Plea: Given the limited human and financial resources
available to him, the author would like to invite interested and qualified
parties (such as individual researchers, programming languages research
groups, software-development companies and corporations, research
funding agencies, and other interested parties) to join him in speeding
up the development of the order-/cat\-egory-theo\-retic approach
to modeling Java generics.}}

\bibliographystyle{plain}

\end{document}